\documentclass[twoside,12pt]{article}
\usepackage{epsf,epsfig,float,amssymb,latexsym,amsmath,amsthm,fancyhdr}
\usepackage{graphics,psfrag,longtable}

\newcommand{\be}{\begin{equation}}
\newcommand{\ee}{\end{equation}}
\newcommand{\bea}{\begin{eqnarray}}
\newcommand{\eea}{\end{eqnarray}}

\newcommand{\sigmaterm}{\sigma_{\pi N}}
\begin{document}

%\title{ \vspace{1cm} Relativistic chiral representation of the $\pi N$ scattering amplitude}

\title{ \vspace{1cm} {\Huge Relativistic chiral representation of the $\pi N$ scattering amplitude I:}\\ {\LARGE The Goldberger-Treiman relation}}

\author{J.~M.~Alarc\'on$^1$, J.~Mart\'{\i}n~Camalich$^{2}$ and J.~A. Oller$^1$\\
\\
$^1${\it {\small Departamento de F\'{\i}sica. Universidad de Murcia. E-30071,
Murcia. Spain}}\\
$^2${\it {\small Department of Physics and Astronomy, University of Sussex, BN1 9QH, Brighton, UK}}}
\maketitle
\begin{abstract}
In this work we study the $\pi N$ scattering process within the Baryon Chiral Perturbation Theory framework in the covariant scheme of Extended-On-Mass-Shell (EOMS). 
We compare the description obtained in this scheme with the previously obtained using the Infrared Regularization scheme and show that EOMS accomplishes the best convergence, being 
able to  extract from partial wave analyses reliable values of important quantities as the Goldberger-Treiman deviation. In regard to the latter, we solve 
the long-standing problem concerning to the extraction of the Goldberger-Treiman deviation with covariant ChPT that jeopardized the applicability of ChPT to the $\pi N$ system.
We also show the potential of the unitarization techniques applied to the perturbative calculation in the EOMS scheme, that allow us to increase the range of validity of our description 
up to $\approx 200$~MeV in $\sqrt{s}$.
\end{abstract}
%\eject
%\tableofcontents
%-----------------------------------------------------------------------------

 \section{Introduction}

The $\pi N$ scattering is a process thoroughly studied experimentally and, in fact, we have experimental data since sixties. From the theoretical point of view, it is the basic hadronic process involving baryons and one of 
the most important test ground for Chiral Perturbation Theory with Baryons (BChPT). The first attempt to study this process using BChPT was performed by Gasser, Sainio and Svarc in their seminal work \cite{G&S&S} 
using a fully covariant approach. In this work, they realized that when one deals with nucleons, a new heavy scale appears in the ChPT formalism that does {\itshape not vanishes in the chiral limit} and spoils the 
standard power counting of ChPT. In order to solve this problem, Jenkins and Manohar invented the Heavy Baryon Chiral Perturbation Theory approach \cite{J&M} HBChPT, which integrates
 out the heavy degrees of freedom of the nucleon expanding the Lagrangian in series $1/m_N$, with $m_N$ the nucleon mass. This formalism describes well the low-energy physical region \cite{F&M&S, F&M-Op4, F&M-Delta, B&M} at 
the cost of losing Lorentz invariance, although it does not converge in the subthreshold region \cite{B&K&M-Int.J.Mod.Phys, B&L}
. This means that we cannot check some chiral symmetry predictions for QCD (low energy theorems). In order to solve this problem of convergence, Becher and Leutwyler invented the Infrared Regularization scheme (IR) \cite{IR},
that recovers the standard power counting of ChPT keeping manifest Lorentz invariance. This improves the convergence with respect to HBChPT and converges in the subthreshold region. However, as was show in \cite{B&L}, the one-loop 
representation is not precise enough to allow an accurate extrapolation of the physical data to the Cheng-Dashen point to extract the value of $\sigmaterm$. Later works showed that the IR 
description of the phase shifts are of the same quality as those of HBChPT \cite{nuestroIR}, although a {\itshape huge} and {\itshape strongly scale dependent} Goldberger-Treiman (GT) relation deviation is found in this scheme \cite{T&E, nuestroIR}.
In fact, the scale dependence is one of the main characteristics of IR. Another important limitation of IR comes out when we 
reach energies that can make the Mandelstam variable $u=0$ (this energy corresponds to $\sqrt{s}\gtrsim 1.34$~GeV for $\pi N$ scattering), because in this case 
the amplitudes develop an unphysical cut that limits the high energy description and, therefore, the applicability of Unitarization methods \cite{nuestroIR}. This unphysical cut is also responsible for a bad prediction 
for the magnetic moments when using this scheme \cite{MomentosMagneticos}. In order to overcome the problems that one encounters in the IR scheme keeping the good analytical properties of a covariant approach, we 
calculated de $\pi N$ scattering amplitude in ChPT up to $\mathcal{O}(p^3)$ in the chiral expansion, using the so-called {\itshape Extended-On-Mass-Shell} scheme (EOMS) \cite{F&G&J&S}. This scheme recovers the spirit of 
the full covariant approach performed by \cite{G&S&S} but keeping the standard power counting of ChPT through a renormalization of the low energy constants (LECs) that appear in the Lagrangian. So, the EOMS scheme 
can be considered as a second renormalization in the sense that the first renormalization would be the $\overline{MS}$ renormalization that cancel the infinities that come from the loop diagrams, and the EOMS renormalization 
is the renormalization that cancel the power counting breaking terms (PCBT) that appear in the covariant approach. The proof that this renormalization can cancel {\itshape all} the PCBT comes form the IR formalism because 
Becher and Leutwyler proved that all the PCBT are contained in what they called the {\itshape regular part} of the loop integral, that is analytical in the quark masses and momenta, so 
that means that can be absorbed in the most general Lagrangian. The advantages of this scheme over the IR one are \cite{nuestroEOMS}: 1) we do {\itshape not} have to deal with any scale dependence, 
2) the contribution of the loop diagrams to the GT deviation is very small ($\approx 0.2\%$) that is of the size of what we would expect from explicit symmetry breaking, 
3) our amplitudes are free from unphysical cuts, what means that they have the right analytical properties in the whole energy plane. 
In this proceeding we will focus on the comparison between both covariant methods: EOMS and IR.

%-------------------------------------------------------------------------

\section{Perturbative Calculations}

\begin{figure}%[ht]
\begin{center}
\psfrag{ss}{{\tiny $\sqrt{s}$ (GeV)}}
\psfrag{S11per}{{\footnotesize $S_{11}$}}
\psfrag{S31per}{{\footnotesize $S_{31}$}}
\psfrag{P11per}{{\footnotesize $P_{11}$}}
\psfrag{P13per}{{\footnotesize $P_{13}$}}
\psfrag{P31per}{{\footnotesize $P_{31}$}}
\psfrag{P33per}{{\footnotesize $P_{33}$}}
\epsfig{file=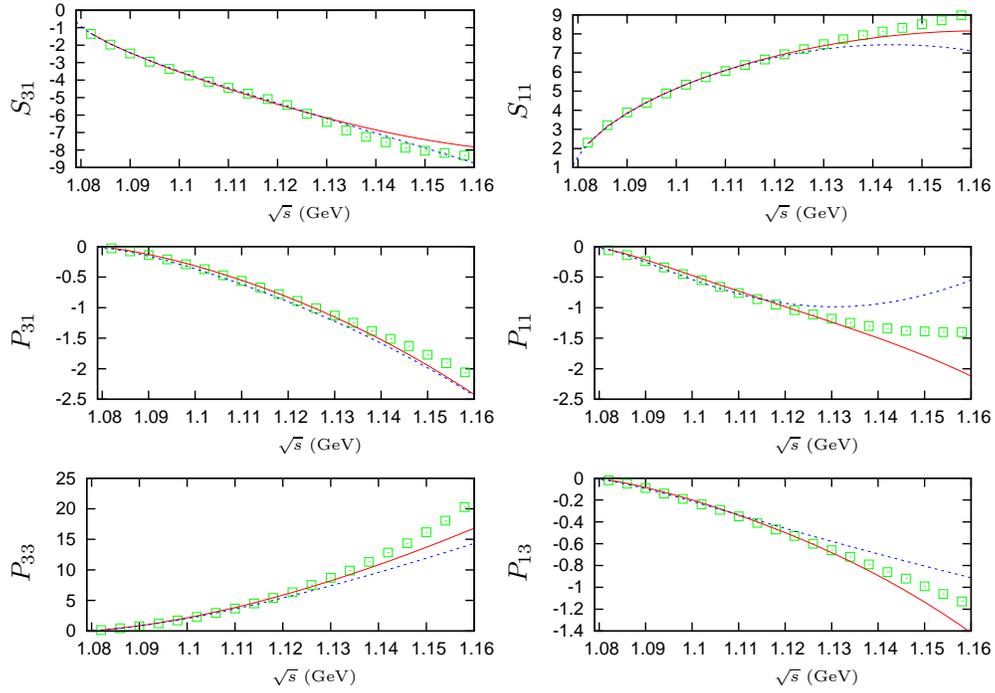,width=.50\textwidth,angle=-90}
\caption[pilf]{\protect \small Fits to KA85 \cite{KA85}. Solid line corresponds to the EOMS result and the dashed to IR. Both fits are performed up to $\sqrt{s}_{max}=1.13$~GeV \label{KA85-pert}}
\end{center}
\end{figure}

\begin{figure}%[ht]
\begin{center}
\psfrag{ss}{{\tiny $\sqrt{s}$ (GeV)}}
\psfrag{S11per}{{\footnotesize $S_{11}$}}
\psfrag{S31per}{{\footnotesize $S_{31}$}}
\psfrag{P11per}{{\footnotesize $P_{11}$}}
\psfrag{P13per}{{\footnotesize $P_{13}$}}
\psfrag{P31per}{{\footnotesize $P_{31}$}}
\psfrag{P33per}{{\footnotesize $P_{33}$}}
\epsfig{file=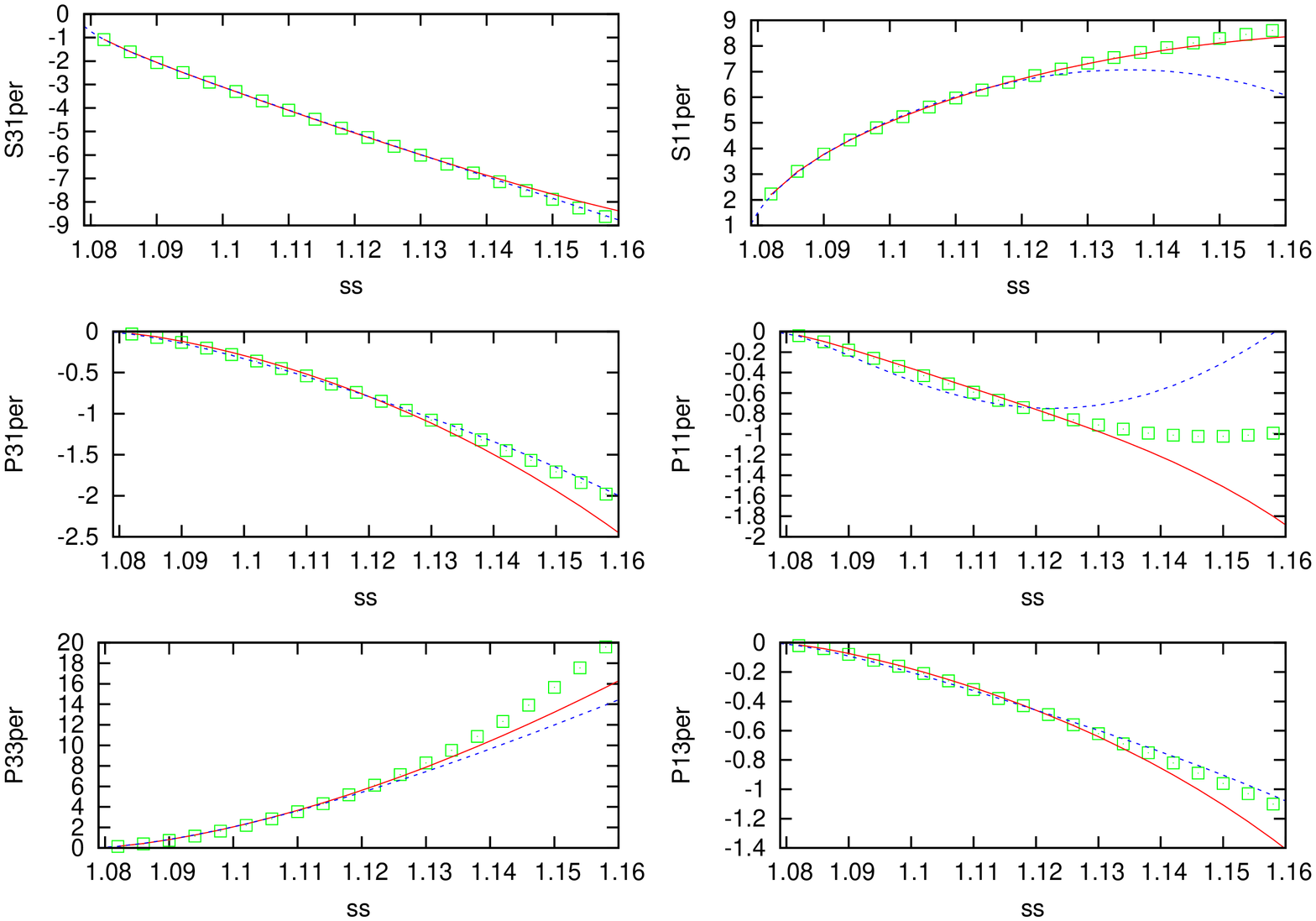,width=.50\textwidth,angle=-90}
\caption[pilf]{\protect \small Fits to WI08 \cite{WI08}. Solid line corresponds to the EOMS result and the dashed to IR. Both fits are performed up to $\sqrt{s}_{max}=1.13$~GeV \label{WI08-pert}}
\end{center}
\end{figure}

To compare fairly both methods we proceed with EOMS in the same way as we did with IR, so we perform the perturbative study as in \cite{nuestroIR}. We considered the partial wave analyses 
(PWAs) of the Karlsruhe group \cite{KA85} (KA85) and the current solution of the George Washington University group \cite{WI08} (WI08), assigning the same errors as we did 
in \cite{nuestroIR}. The results of the fits are shown in Fig. \ref{KA85-pert} and Fig. \ref{WI08-pert}

\begin{table}%[ht]
 \begin{center}
\begin{tabular}{|r|r|r|r|r|r|}
\hline
\small{LEC}       & KA85-EOMS	        &    WI08-EOMS        & KA85-IR 			   & WI08-IR			     	     & HBCHPT			       \\
                  & ${\cal O}(p^3)$     & ${\cal O}(p^3)$     & ${\cal O}(p^3)$ \cite{nuestroIR}   & ${\cal O}(p^3)$ \cite{nuestroIR}& ${\cal O}(p^3)$  \cite{F&M&S}  \\
\hline							   	    								     				
$c_1$              &  $-1.26\pm 0.07$   &  $-1.50\pm 0.06$ & $ -0.71\pm 0.49$			   & $-0.27 \pm 0.51$		     & $(-1.71,-1.07)$  		\\   
$c_2$              &  $ 4.08 \pm 0.09$  &  $3.74\pm 0.09$  & $ 4.32\pm 0.27$			   & $4.28 \pm 0.27$		     & $(3.0,3.5)$			\\
$c_3$              &  $-6.74 \pm 0.08$  &  $-6.63\pm 0.08$ & $ -6.53\pm 0.33$ 		           & $-6.76 \pm 0.27$		     & $(-6.3,-5.8)$			\\
$c_4$              &  $3.74\pm 0.05$    &  $3.68\pm 0.05$  & $ 3.87\pm 0.15$                       & $ 4.08\pm 0.13$		     & $(3.4,3.6)$			      \\
\hline		  					   				    					     				
$d_1+d_2$          &  $3.25\pm 0.55$    &  $3.67\pm 0.54$  & $ 2.48\pm0.59 $			   &  $ 2.53\pm 0.60$		     & $(3.2,4.1)$			   \\
$d_3$              & $-2.72 \pm 0.51$   &  $-2.63\pm 0.51$ & $-2.68\pm 1.02$			   & $-3.65 \pm 1.01$		     & $(-4.3,-2.6)$			       \\
$d_5$              & $0.50 \pm 0.13$    &  $-0.07\pm 0.13$ & $2.69 \pm 2.20$		           & $ 5.38\pm 2.40$		      & $(-1.1,0.4)$			     \\
$d_{14}-d_{15}$    & $-6.10\pm 1.08$    &  $-6.80\pm 1.07$ & $-1.71 \pm 0.73$  		           & $ -1.17\pm 1.00$		      & $(-5.1,-4.3)$			       \\
$d_{18}$           & $-2.96\pm 1.44$    &  $-0.50\pm 1.43$ & $ -0.26\pm 0.40$ 		           & $ -0.86\pm 0.43$		      & $(-1.6,-0.5)$				    \\ 
\hline
$\chi^2_{d.o.f.}$  &       $0.35$       &     $0.22$       & $\lesssim 1$			  & $\lesssim 1$  		      &     -					     \\
\hline
$\Delta_{GT}$        &   $ 9\pm 4  \%$  &    $2 \pm  4 \%$ &   $ (20-30\%)$			  &   $ (20-30\%)$		      &    (input)					 \\
\hline    
%$g_{\pi N}$        &   $ 14.03(52)$  &    $ 13.13(52)$     	   &   -		          &   -			      &    -							       \\
%\hline
\end{tabular}
\caption[pilf]{Comparison between LECs and the resulting $\Delta_{GT}$ in the different approaches of BChPT. \label{LECs}}
%{\caption[pilf]{\protect \small Fitted LECs in units of GeV$^{-1}$ ($c_i$) and GeV$^{-2}$ ($d_i$) for the fits WI08-1 and WI08-2 with $\sqrt{s}_{max}=1.13$~GeV. The last columns 
%is the average of all the fits in Tables~\ref{table.cs.ds.pert.ka85} and \ref{table.ir.pert.cs.ds.wi08}.
% \label{LECs}}}
\end{center}
\end{table}

The fitted values of the LECs are given in Table \ref{LECs}. In this table one can compare the EOMS results with those of the IR and HBChPT methods. One can see that the EOMS results for the LECs 
are compatible with both approaches being able to give a better description than IR (lower $\chi^2_{d.o.f.}$). At this point it is important to stress two things: First,
we see in Fig. \ref{KA85-pert} and Fig. \ref{WI08-pert} that with the EOMS scheme we do not have the problems that we encountered in \cite{nuestroIR} when we tried to fit the $P_{11}$ phase shift of WI08 with 
the IR scheme (dashed line Fig. \ref{WI08-pert}). Second, and more important, with the EOMS scheme we solved the problem of the huge $\Delta_{GT}$ of the IR scheme 
that jeopardized the applicability of ChPT to the $\pi N$ system. Within the EOMS scheme we can {\itshape extract from data} values for $\Delta_{GT}$ compatible with the ones reported by the corresponding PWAs. Although the values presented in Table \ref{LECs} for the EOMS result may be consider not very accurate and quite large for KA85-EOMS, it is important to stress that these 
results can be considerably improved once we include explicitly the $\Delta(1232)$ in our EOMS calculation \cite{nuestrosigmaterm}. In this case we obtain very accurate predictions for $\Delta_{GT}$ that are 
perfectly compatible with their corresponding PWAs \cite{nuestrosigmaterm}.  
For a deeper understanding of what is happening in the covariant methods, we will explain briefly the method used for the extraction of the GT deviation. As we did in \cite{nuestroIR}, 
the method consist in taking the limit :

\begin{align*}
 \lim_{s\rightarrow m_N^2} \frac{T^{\mathcal{O}(p^3)}}{T^{\mathcal{O}(p)}}=\left(\frac{g_{\pi N}}{g_A m_N/f_\pi}\right)^2=(1+\Delta_{GT})^2
\end{align*}
 
Where $T^{\mathcal{O}(p^3)}$ and $T^{\mathcal{O}(p)}$ mean the full amplitude calculated up to $\mathcal{O}(p^3)$ and $\mathcal{O}(p)$ respectively. In ChPT $\Delta_{GT}$ is directly 
related to the LEC $d_{18}$ plus a higher order contribution due to the loops ($\Delta_{loops}$):

\begin{align*}
 \Delta_{GT}=-\frac{2 M_\pi^2 d_{18}}{g_A}+\Delta_{loops}
\end{align*}

From fits to PWAs one obtains a natural value for $d_{18}$ from both covariant schemes (EOMS and IR), but when one calculates explicitly the value of $\Delta_{loops}$ it turns out 
that the IR scheme gives huge values that depend strongly on the renormalization scale ($20-30\%$) \cite{nuestroIR} while EOMS gives a {\itshape scale-independent value that is of 
the size that we would expect form explicit chiral symmetry breaking} ($\approx 0.2\%$), solving the long standing problem that covariant BChPT had with this observable. This means 
that the huge $\Delta_{GT}$ is due to the IR prescription, not to a problem of BChPT.

%%%%%%%%%%%%%%%%%%%%%%%%%%%%%%%%%%%%%%%%%%%%%%

\section{Unitarized Calculations}

Another limitation that we encountered in the IR scheme concerns to the applicability of Unitarization techniques to the perturbative calculation. In IR we found that Unitarization 
techniques are limited up to $\sqrt{s}\approx 1.25$~GeV due to the unphysical cut that this scheme introduces \cite{nuestroIR}. So, it is interesting to study if a covariant calculation 
without this unphysical cut could improve the description of the phase shifts. With this aim we implement unitarity to the EOMS-BChPT $\pi N$ amplitude 
and take care of the analyticity properties associated with the right-hand cut writing our unitarized amplitude $T_{IJ\ell}$ by means of an interaction kernel 
$\mathcal{T}_{IJ\ell}$ and the unitary pion-nucleon loop function $g(s)$:   

\begin{align*}
T_{IJ\ell}=\frac{1}{\mathcal{T}^{-1}_{IJ\ell}+g(s)}
\end{align*}

Where $T_{IJ\ell}$ is the amplitude with definite isospin $I$, total angular momentum $J$ and orbital angular momentum $\ell$. Written in this form, $T_{IJ\ell}$ satisfies unitarity 
{\itshape exactly} and the interaction kernel $\mathcal{T}_{IJ\ell}$ can be obtained by matching order by order with the perturbative result \cite{O&M}. On the other hand, the 
subtraction constant $a_1$ contained in the unitary pion-nucleon loop function $g(s)$ is fixed by requiring that $g$ vanishes in the nucleon pole $g(m_N^2)=0$ so in this point 
we recover the perturbative calculation and keep the $P_{11}$ nucleon pole in its right position. On the other hand, in order to take into account the contribution of the 
$\Delta(1232)$ we introduce, in the $P_{33}$ partial wave, a Castillejo-Dalitz-Dyson pole (CDD) \cite{C&D&D} that is a pole that conserves the discontinuities of the partial wave amplitude 
across the cuts. For this partial wave, the unitarized amplitude reads:  $T_{\frac{3}{2}\frac{3}{2}1}=\left(\mathcal{T}_{\frac{3}{2}\frac{3}{2}1}^{-1}+\frac{\gamma}{s-s_P}+g(s)\right)^{-1}$, where the CDD corresponds to $\frac{\gamma}{s-s_P}$, and gives rise to a zero at $s_P$ in $T_{\frac{3}{2}\frac{3}{2}1}$.

\begin{figure}%[ht]
\begin{center}
\psfrag{ss}{{\tiny $\sqrt{s}$ (GeV)}}
\psfrag{S11}{{\footnotesize $S_{11}$}}
\psfrag{S31}{{\footnotesize $S_{31}$}}
\psfrag{P11}{{\footnotesize $P_{11}$}}
\psfrag{P13}{{\footnotesize $P_{13}$}}
\psfrag{P31}{{\footnotesize $P_{31}$}}
\psfrag{P33}{{\footnotesize $P_{33}$}}
\hspace{-0.5cm}
 \epsfig{file=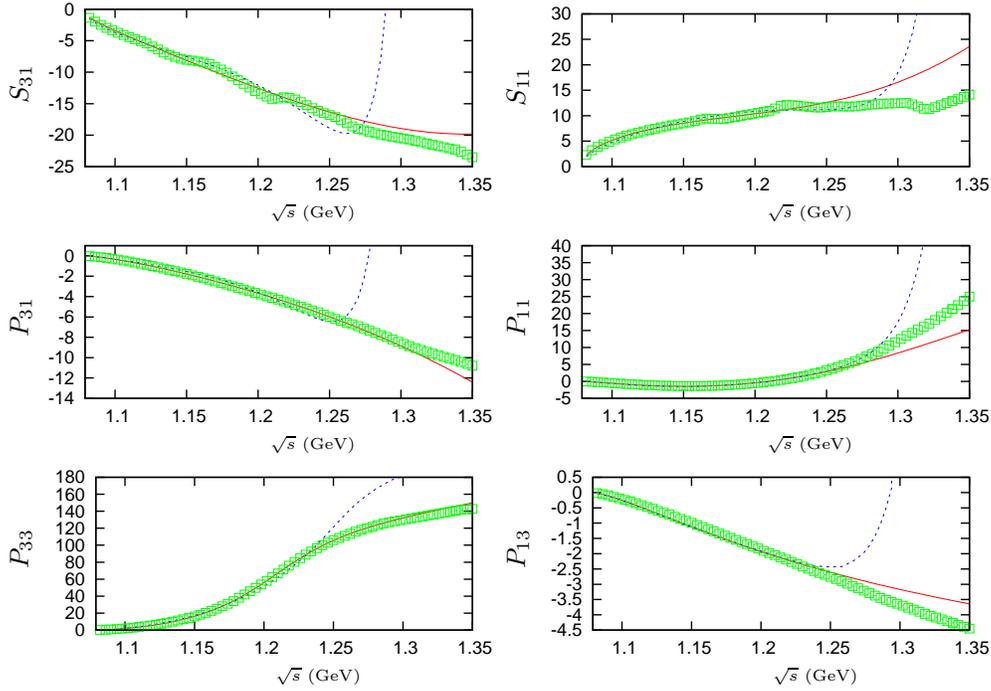,width=.50\textwidth,angle=-90}
\caption[pilf]{Unitarized fits to KA85. Solid line: EOMS. Dashed line: IR \cite{nuestroIR} \label{KA85-uni}} 
\end{center}
\end{figure}

\begin{figure}%[ht]
\begin{center}
\psfrag{ss}{{\tiny $\sqrt{s}$ (GeV)}}
\psfrag{S11}{{\footnotesize $S_{11}$}}
\psfrag{S31}{{\footnotesize $S_{31}$}}
\psfrag{P11}{{\footnotesize $P_{11}$}}
\psfrag{P13}{{\footnotesize $P_{13}$}}
\psfrag{P31}{{\footnotesize $P_{31}$}}
\psfrag{P33}{{\footnotesize $P_{33}$}}
\hspace{-0.5cm}
 \epsfig{file=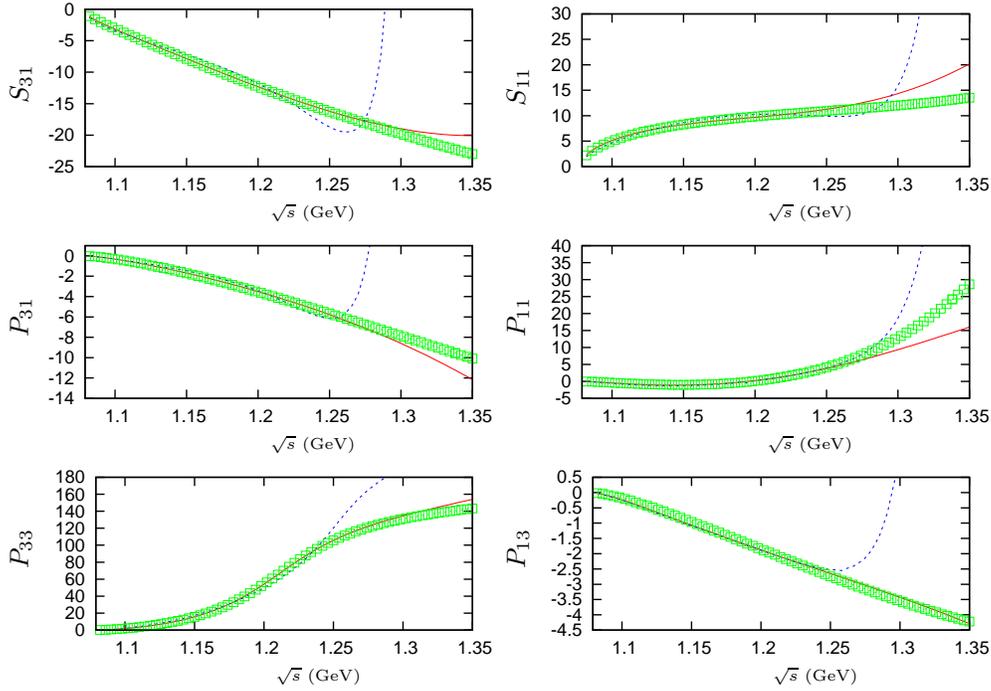,width=.50\textwidth,angle=-90} 
\caption[pilf]{Unitarized fits to WI08. Solid line: EOMS. Dashed line: IR \cite{nuestroIR} \label{WI08-uni}} 
\end{center}
\end{figure} 

In both Fig. \ref{KA85-uni} and Fig. \ref{WI08-uni} one can see how the kernel calculated in the EOMS scheme provides a good description of the phase shifts up to 
$\sqrt{s}\approx 1.35$~GeV while IR can only reproduce them up to $\sqrt{s}\approx 1.25$~GeV due to the unphysical cut that this scheme introduces. 
This corresponds to an increase of $\approx100$~MeV in $\sqrt{s}$ compared to IR. On the other hand the CDD is able to reproduce perfectly the raise of the $P_{33}$ 
phase shift due to the $\Delta(1232)$, and if one compares the description of the unitarized and the perturbative
 amplitudes, one finds a drastic increase in the energy region of the data ($\approx 200$~MeV in $\sqrt{s}$).

%%%%%%%%%%%%%%%%%%%%%%%%%%%%%%%%%%%%%%%%%

%%%%%%%%%%%%%%%%%%%%%%%%%%%%%%%%%

\section{Summary and Conclusions}

In summary, the $\pi N$ scattering is a fundamental process that provides the basic test for ChPT with baryons. There have been many attempts to describe this process in BChPT 
but every approach has had their own problems: {\itshape lack of convergence, unphysical cuts, unphysically large GT deviation}, etc. These problems questioned the applicability of 
ChPT to the $\pi N$ system. In this work we showed that BChPT in the EOMS scheme solves these issues providing a chiral representation that converges, and giving rise to a GT violation in good agreement with phenomenology.
This and other important quantities such as the $\sigmaterm$ can be extracted form PWAs accurately and reliably once we introduce explicitly the contribution 
of the $\Delta(1232)$ in our calculations, as we showed in \cite{nuestrosigmaterm}, and that part will be explained in \cite{proceedingJorge}.
It is also very interesting to show the potential of the Unitarization techniques applied to a kernel with good analytical properties. In this work we show that with this unitarization 
method we could increase the range of 
our description up to $\sqrt{s}\approx 1.35$~GeV, that means an improvement of $\approx 200$~MeV in $\sqrt{s}$ with respect to the perturbative calculation. 
Compared with IR, the unitarized EOMS amplitudes achieve a good description up to energies $\approx100$~MeV higher in $\sqrt{s}$.

%%%%%%%%%%%%%%%%%%%%%%%%%%%%%%%%%%%%%%%%%%%%%%%%%%

%%%%%%%%%%%%%%%%%%%%%%%%%%%%%%%%%%%%%%%%%%%%%%%%%%%%%%%
%%%%%%%%%%%%%%%%%%%%%%%%%%%%%%%%%%%%%%%%%%%%%%%%%%%%%%%%%%%
%%%%%%%%%%%%%%%%%%%%%%%%%%%%%%%%%%%%
%%%%%%%%%%%%%%%%%%%%%%%%%%%%%%%%%%%%%%%%%%%%%%%%%%%%%%%%%%%%%%%%%%%%
%%%%%%%%%%%%%%%%%%%%%%%%%%%%%%%%%%%%%%%%%%%%%%%%%%%%%%

%\newpage

  \end{document}